\documentclass[twocolumn]{elsart}
\usepackage{amssymb}
\usepackage{graphicx}
\begin{document}

\begin{frontmatter}
\title{Security of Y-00 and similar quantum cryptographic protocols}
\author{Horace P. Yuen}\ead{yuen@ece.northwestern.edu},
\author{Prem Kumar},
\author{Eric Corndorf},
\author{and Ranjith Nair}
\address{Center for Photonic Communication and Computing\\
Department of Electrical and Computer Engineering\\
Department of Physics and Astronomy\\
Northwestern University, Evanston, IL 60208}

\begin{abstract}
It is claimed in Phys.\ Lett.\ A by T. Nishioka {\it et al}.
[{\bf 327} (2004) 28--32] that the security of Y-00 is equivalent
to that of a classical stream cipher. In this paper it is shown
that the claim is false in either the use of Y-00 for direct
encryption or key generation, in all the parameter ranges it is supposed
to operate including those of the experiments reported thus far.
The security of Y-00 type protocols is clarified.
\end{abstract}
\end{frontmatter}
A new approach to quantum cryptography called KCQ, (keyed
communication in quantum noise), has been developed \cite{yuen04}
on the basis of a different advantage creation principle from that
in either uncorrelated-classical-noise key generation
\cite{maurer93} or the well known BB84 quantum protocol
\cite{bennett84}.  A special case called $\alpha\eta$ (or Y-00 in
Japan) has been experimentally investigated and developed to a
considerable extent \cite{barbosa_1,corndorf_1,corndorf_2,corndorf_3,corndorf_4} for direct encryption.
In Ref.\cite{Nishioka}, the claim is made that Y-00 is equivalent
to a classical stream cipher, in particular that the quantum noise is negligible, and thus also cannot be used for key generation.
 This claim is justified by an "attack" that reduces the security of Y-00 to that of a standard stream cipher for the
 purpose of obtaining the data bits from observing the output of Y-00. In this paper, we will show that this claim is
 patently false.

The main explicit claim in \cite{Nishioka} is that their classical stream cipher, ``Case 2'', has the same security as Y-00,
 and so can be employed instead. We will refute this claim in connection with both data and key security
(the latter is
not even considered in \cite{Nishioka})
, in direct
encryption as well as in key generation, and also show that their ``attack'' is an ineffective one on Y-00.

One basic error in \cite{Nishioka} is the assumption that Y-00 with the parameters reported in \cite{barbosa_1,corndorf_1,corndorf_2,corndorf_3,corndorf_4}
is reducible to their ``Case 1'' cipher for which Eq. (10) of \cite{Nishioka} is valid without error. Such error of course decreases
with increasing coherent-state energy, but
it is trivial to claim that a coherent-state system is classical when
the energy in the system is large enough as compared to all the parameters of the operating scheme.
We have always qualified our own claim by saying that the coherent-state energy is``mesoscopic''. In the case of direct encryption
parameters reported experimentally \cite{barbosa_1,corndorf_1,corndorf_2,corndorf_3,corndorf_4}, the reduction of Ref. \cite{Nishioka}
results in a classical stream cipher \textit{in quantum noise} with an error rate of $\sim$1\%
, and has already been analyzed in detail by
the Hirota group \cite{hirota03}. Furthermore, even when the coherent-state quantum noise of Y-00 can in principle be
replaced by classical randomization, such randomization makes Y-00 a random cipher. It is known that a random cipher
 may have better secret-key
security compared to a classical stream cipher, such as ``Case 2'' of \cite{Nishioka}, which is nonrandom.

Another error
made in \cite{Nishioka} may arise from the incorrect
claim made in \cite{barbosa_2}. This involves Fig. 4 of \cite{Nishioka} and the discussion
around it pertaining
to the use of Y-00 for key generation, with the key being used subsequently in a classical cipher. The protocol of Fig. 4 is seriously
incomplete for key generation and is not one we intended or claimed to use.
Before further elaboration on
these errors in \cite{Nishioka}, we first briefly review the Y-00 scheme and remove a very common misconception about direct encryption versus
key generation.

\begin{figure*}\begin{center}
\includegraphics[width=13cm]{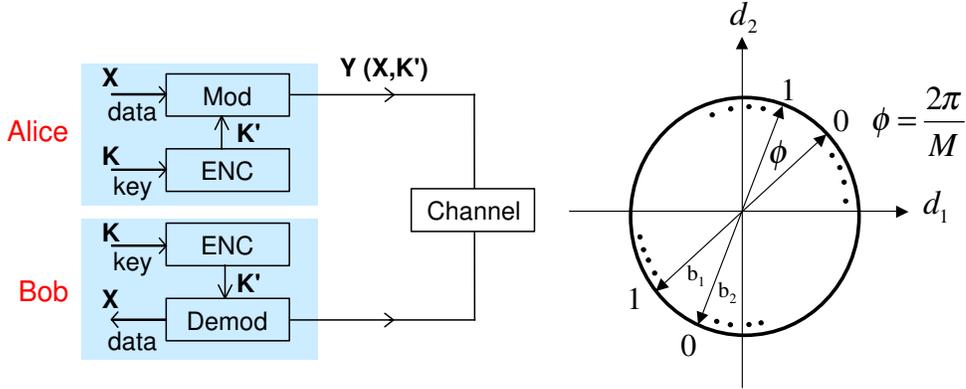}\caption{Left: Overall schematic of the Y-00 scheme.  Right: Depiction of $M/2$
bases with interleaved logical state mappings.}\label{setup}
\end{center}\end{figure*}

Consider the original experimental scheme Y-00 as
described in Ref.~\cite{barbosa_1} and depicted in Fig.~\ref{setup}. Alice
encodes each data bit into a coherent state in a \textit{qumode},
an infinite-dimensional Hilbert space, of the form \cite{note}
\begin{equation} \label{states}
    |\alpha_{\ell}\rangle=|\alpha_{0}(\cos\theta_{\ell}+i
    \sin\theta_{\ell})\rangle
\end{equation}
where $\alpha_{0}$ is real, $\theta_{\ell}=2\pi\ell/M$, and $\ell \in \{0,..., m-1\}$. The $M$
states are divided into $M/2$ basis pairs of antipodal
signals $\{|\pm \alpha_{\ell} \rangle \}$ with
$-\alpha_{\ell}=\alpha_{\ell+M/2}$.  A seed key $K$ of bit length
$|K|$ is used to drive a conventional encryption mechanism whose
output is a much longer running key $K^{\prime}$ that is used to
determine, for each qumode carrying the bit $b \{=0,1\}$, which
pair $\{|\pm\alpha_{\ell}\rangle\}$ is to be used. Bob utilizes a
quantum receiver to decide on $b$ knowing which particular pair
$\{|\pm\alpha_{\ell}\rangle\}$ is to be discriminated. On the
other hand, Eve needs to pick a quantum measurement for her attack
in the absence of the basis knowledge provided by the seed or
running key.  The difference in the resulting receiver performance
is a quantum effect with \textit{no} classical analog, and
constitutes the ground for possible advantage creation in the
scheme. Note that since the quantum-measurement noise is irreducible, such advantage creation can result in
an unconditionally secure key generation protocol.
In contrast, in a classical situation including noise, the simultaneous
measurement of the amplitude and phase of the signal, as realized
optically by heterodyning, provides the general optimal
measurement for both Bob and Eve; thus preventing any advantage
creation under our approach that grants Eve a copy of the state for the purpose of bounding her information.

One needs to first distinguish the use of such a scheme for key generation versus data
encryption.  It may first appear that if the system is secure for
data encryption, it would also be secure for key generation if the
data are subsequently used as keys. This is indeed the view taken in Ref.~\cite{Nishioka} and Ref.~\cite{barbosa_2}.
It is unfortunate that the author of Ref.~\cite{barbosa_2}, a co-author of Refs.\ \cite{barbosa_1,corndorf_1}, made this conclusion
that the direct encryption experiments in \cite{barbosa_1,corndorf_1} would
already
allow key generation inspite of our objections. In fact, for the direct encryption experiments in
Refs.~\cite{barbosa_1,corndorf_1,corndorf_2,corndorf_3,corndorf_4}, we
have only claimed \emph{complexity-based} security against
general attacks, with ``unconditional security'' only against a
very limited class of ``individual attacks.''
The situation may be delineated as follows.  Let $X_{n},Y_{n}^{\rm E},Y_{n}^{\rm B}$
be the classical random vectors describing the bit data of length $n$, Eve's
observation, and Bob's observation. Eve may
make any quantum measurement on her copy of the quantum signal to obtain $Y^{\rm E}_{n}$ in her attack.
In the case of a standard classical cipher,
$Y^{\rm E}_{n}=Y_{n}^{\rm B}=Y_{n}$, the following Shannon limit \cite{shannon49} applies
\begin{equation} \label{shannonlimit}
H(X_n|Y_n) \leq H(K)
\end{equation}
and so there can be no \textit{fresh} key generation.  This is
because all the uncertainty in $X_{n}$ is derived from $K$,
however long $n$ is. While $H(X_n|Y_n^E)$ describes the level of information-theoretic security of the data $X_n$ against
ciphertext-only attacks, $H(K|Y_n^E)$ describes the information-theoretic security of the key against ciphertext-only attacks with
known a priori probability $p(X_n)$, thus including known and chosen plaintext attacks in the case of degenerate $p(X_n)$. See Ref.
\cite{acm} and \cite{yuen04}  for further discussion. In standard cryptography, one typically does not worry about ciphertext-only
attack on completely random data, where Eq.~(\ref{shannonlimit}) is usually satisfied with equality for large $n$ for the designed key
length $ |K|=H(K)$. Rather, it is attacks on the key with known nonuniform $p(X_n)$, using information on $K$ so obtained on
future data, that is the focus of concern, as in the Advanced Encryption Standard (AES).

The reduction of Y-00 to the classical stream cipher of ref \cite{Nishioka} consists in collapsing any observation to a single
bit $l_i$, with the claim that, as described in Eq.~(9)-(11) of \cite{Nishioka},

\begin{equation} \label{nishioka}
l_i=x_i \oplus \tilde{k_i}
\end{equation}

where $x_i$ is the data bit \cite{note2} at the $i$th position of the data sequence, and $\tilde{k_i}$ is a fixed function of the running key that
determines the basis used for that position. Each $l_i$ is $0$ or $1$ according to whether Eve's observation on the $i$th qumode lies on the
upper or lower half-circle with respect to the ``horizontal'' basis given by the all zero running key. However, Eq.~(\ref{nishioka}) is not
always true
due to the
quantum noise in Eve's measurement which sometimes pushes the measured result to the wrong side of the horizontal line.
From the intrinsic coherent-state angular uncertainty with a phase standard deviation of $1/{\alpha_0}$, an estimate of the bit error
rate
\begin{equation} \label{ber}
P_b^E \sim 2/(\pi\alpha_0)
\end{equation}
is simply obtained if one assumes that the measured state is uniformly distributed within a standard
deviation only. In deriving Eq.~(\ref{ber}) we have also used the fact that the $(M/2)$ bases are selected with uniform marginal probability
for each qumode, which is a consequence of using, e.g., a LFSR for the ENC box of Fig. 1 with seed key length $|K| \geq log_{2}(M/2)$.
This $P_b^E$ is in rough agreement with the numerical calculations of ref \cite{hirota03}, which includes the optimal quantum
receiver performance result for this ``attack'' via the optimal binary decision measurement. The resulting 1\% error means that for the
purpose of attacking the data $X_n$, the reduction is equivalent to a classical stream cipher with unknown $K$ received in noise that
causes 1\% error in the output ciphertext. Thus, Y-00 is \textit{not} equivalent to a classical stream cipher, but rather to one
in significant noise
even in the experimental regime reported thus far. Indeed, not only do such errors invalidate the Shannon limit Eq.~(\ref{shannonlimit})
for a standard stream cipher,
they also create advantage for the users and allow key generation in the usual fashion \cite{maurer93}.
To see that the error rate of 1\% is significant, note that it allows a substantial key generation rate of ~10 Mbps for a raw bit rate
of ~1 Gbps, using privacy amplification \cite{bennett95}. The authors of \cite{Nishioka} mistakenly omit the privacy
amplification step required for key generation in their Fig. 4.

On the other hand, a stronger attack may be launched on Y-00 by making a heterodyne measurement which
retains all the $log_2 M$ bits of output
for each qumode. Under such an attack, the cipher becomes a classical random cipher in principle, satisfying Eq.~(\ref{shannonlimit}) with
the experimental
parameters of \cite{barbosa_1}. This is because the experiments on the original Y-00 have parameters that satisfy

\begin{equation} \label{eve}
H(X_n|Y_n^E, K) \sim 0
\end{equation}

when the heterodyne measurement is made on each qumode by Eve. Under Eq.~(\ref{eve}), Eq.~(\ref{shannonlimit}) also obtains and the data security
is no better than $|K|$ as in all standard symmetric key ciphers.
Furthermore, in this regime, and under the heterodyne attack which is more powerful than that of \cite{Nishioka}, key generation with information-theoretic
security is impossible in principle, a point \textit{missed} in ref \cite{barbosa_2} and in
all the criticisms of Y-00 including ref \cite{Nishioka} and ref \cite{loko}, but was explicitly stated in the first version of
ref \cite{yuen04}. This point is at least implicit in ref \cite{barbosa_1} where we said the experiment has to be modified for key
generation. One simple way to break the Shannon limit Eq.~(\ref{shannonlimit}) while protecting the key at the same time is to randomize
(unkeyed) the state transmitted to cover the half-circle defined by the basis chosen by the running key, which we call DSR in
\cite{yuen04}. Indeed, the resulting noise behavior for Eve is similar to the 1\% error neglected in ref \cite{Nishioka}, and is also
the basis of advantage creation for key generation. Clearly, there is no room to go into any detail on such variations and extensions
of Y-00 in this paper.

Nevertheless, it is important to note that heterodyning by Eve does not reduce Y-00 to a
classical stream cipher even under Eq.~(\ref{eve}). Rather, it
reduces it to a \textit{random cipher}, i.e., a cipher with randomized encryption \cite{massey88} so that

\begin{equation} \label{randomization}
H(Y_n|X_n,K) \neq 0,
\end{equation}

which can be accomplished classically in principle, but not in current practice. This is because true random numbers can only be generated
physically, not by an algorithm, and the practical rate for such generation is five to six orders of magnitude below the $\sim$ Gbps rate
in our experiments where the coherent-state quantum noise does the randomization \textit{automatically}. Furthermore, our physical
``analog'' scheme does not sacrifice bandwidth or data rate compared to other known randomization techniques. There is an unexplored avenue
with respect to a random
cipher in that there is \textit{no proof} that the key is not information-theoretic secure, i.e., that $K$ can be pinned
down by a long $Y_n$ via the unicity distance with
known $p(X_n)$ as in a non-randomized cipher \cite{shannon49,massey88,stinsonbook}, whether $p(X_n)$ is degenerate or not. Indeed,
it is known \cite{jendahl89} that a specific kind of randomized encryption can defeat any attack on the key when the source
generates independent data bits with $p(X=0) \neq 1/2$. Since the coherent-state quantum noise makes efficient high-rate randomized
encryption possible in practice in Y-00, it is indeed a quantum cipher in the important sense that an essential feature of the cipher
arises from quantum noise.

In this connection, we address the attacks described by Lo and Ko \cite{loko}, which can be launched either when a long sequence of
plaintext is known or when the plaintext statistics are nonuniform. Therefore, they are not directly applicable to Y-00 used for key
generation. These attacks can however be launched on a classical cipher that uses the generated key, and the authors of
\cite{loko} give an argument that reduces such an attack to a similar one directly on the data sent in the key generation step.
However, this reduction is incorrect becuase, as in \cite{Nishioka}, the privacy amplification step in the key generation stage is
omitted.  Furthermore, their attacks are impractical in that they require
exponential loss or exponentially long input $n-$sequences \cite{hirota04} and exponential search. They also miss the distinction between random
and non-random ciphers with regard to attacks on the key. Also, the Grover search attack desribed in \cite{loko} is claimed to break Y-00 because in the asymptotic $n \rightarrow
\infty$ limit, the output states corresponding to different seed key values become orthogonal. In addition to the subtle problem of
orthogonality in a non-separable Hilbert space, it makes little cryptographic sense,
even for a non-random cipher, to just look at the asymptotic $n \rightarrow \infty$ limit. Indeed, Shannon calls a system that is broken only at
 $n \rightarrow \infty$ ``ideal'' \cite{shannon49,massey88}.

The claimed ``Case 2'' non-random-cipher reduction of Y-00 in \cite{Nishioka} has weaker security against attacks on the key compared to
Y-00 due to the
1\% error that exists in the attack of \cite{Nishioka} on Y-00. This error induces random
errors on the actual bases or running key estimate, and may allow some information-theoretic security on $K$. Indeed, even under a general
attack, the logical possibility is open that Y-00 is information-theoretic secure or at least Shannon ``ideal''. Even if such turns out
not to be the case,
the ``Case 2'' cipher still has less key security against known-plaintext attacks
than Y-00 for the following reason.
Any given classical nonrandom cipher can be used as the ENC box in Y-00 which
then provides an added layer of protection through the coherent-state modulation. Even under the heterodyne attack that utilizes the
full state observation, one obtains the following brute-force key-search complexity corresponding to the number of possible running
key sequences for large $n$,

\begin{equation} \label{complexity}
C \sim (\frac {\lambda M} {\sqrt{2}\pi \alpha_0})^{|K|/\log_2(\frac {M} {2})}
\end{equation}

where $\lambda=2$ for ciphertext-only attack(i.e. random data) and $\lambda=1$ for known-plaintext attacks. The estimate Eq.~(\ref{complexity})
is obtained by counting only the possible states within one standard deviation of the phase, which is actually an underestimate for large $n$.
With our experimental parameters of
$ M \sim 4 \times 10^3,
\alpha_0 \sim 2 \times 10^2, |K| \sim 4.4 \times 10^3$ \cite{corndorf_4}, one has $C \gtrsim 2^{480}$ for $\lambda=1$,
well beyond any conceivable classical
or quantum search capability. Note that the Grover's search described in \cite{loko} suffers from a similar exponential limitation.
This search is needed
to attack the ENC box seed key from its output, which is absent for a nonrandom classical stream cipher where the ENC output is
\textit{uniquely}
specified in a known-plaintext attack. One may match the ENC cipher rate to the data rate in Y-00 by using a total of
$log_2 \frac {M} {2}$ different deterministic functions $f_i$ to operate on a given running key segment of $log_2 \frac {M} {2}$ bits
to provide the bases for $log_2 \frac {M} {2}$ data bits. Although this would lower the estimate Eq.~(\ref{complexity}) in general,
under a known-plaintext attack a search complexity remains for pinning down the possible outputs of the ENC box whereas
the output of the ENC box is uniquely specified for the ``Case 2''
cipher. Note, however, that for ciphertext-only attacks on $K$ (i.e. those for which the plaintext is random),  a classical stream cipher
can provide information-theoretic security.

We briefly describe the possibility of key generation with the original Y-00 of Fig.1. The condition for
information-theoretically
secure fresh key generation is, in general

\begin{equation} \label{keygeneration}
H(X_n|Y_n^E,K)>H(X_n|Y_n^B,K).
\end{equation}

In Eq.~(\ref{keygeneration}), $Y_n^E$ is obtained from a quantum measurement \textit{without the knowledge of} $K$. It is then used together with any
value of $K$ to estimate the data $X_n$. This necessary condition has to be supplemented with one on the key $K$ security
for defense against adaptive measurements, as discussed in \cite{yuen04}, to make it sufficient also. This would require the extension of
Y-00 in
different possible ways, such as DSR and CPPM described in \cite{yuen04}. However, against individual attacks with a fixed qumode
measurement, Eq.~(\ref{keygeneration}) is sufficient and can be readily seen to hold as follows.  With
$S\equiv|\alpha_0|^2$ being the average photon number in the
states (1), the bit-error rate for Bob with the optimum quantum
receiver \cite{helstrom76} is
\begin{equation}
\label{optimum}
P_{b}=\frac{1}{4}e^{-4S}.
\end{equation}
The bit-error rate for heterodyning, considered as a possible attack, is the well known Gaussian result
\begin{equation}
P_{b}^{\rm het}\sim\frac{1}{2}e^{-S},
\end{equation}
and that for the optimum-phase measurement tailored to the states in (1) is
\begin{equation}
\label{phase}
P_{b}^{\rm ph}\sim\frac{1}{2}e^{-2S}
\end{equation}
over a wide range of $S$.  The difference between Eq.~(9) and Eq.~(11) allows key generation at any value of $S$
if $n$ is long enough. With a mesoscopic signal level $S\sim7$,
one has $P_{b}\sim 10^{-12}$, $P_{b}^{\rm het}\sim 10^{-3}$,
$P_{b}^{\rm ph}\sim 10^{-6}$. For reasonable $n$, this contradicts the claim in \cite{Nishioka} that quantum effects are negligible
until $S < 1+1/\sqrt{2}$, as follows.
If the data arrives at a rate of 1
Gbps, Bob is likely to have $10^{9}$ error-free bits in 1 second,
while Eve would have $\sim 10^{6}$ or $\sim 10^{3}$ errors in her
$10^{9}$ bits with heterodyne or the optimum-phase measurement
(which has no known experimental realization).  With the usual
privacy amplification, the users can then
generate $\sim 10^{6}$ or $\sim 10^{3}$  bits in the 1 second
interval by eliminating Eve's information. While these parameter values are not particularly remarkable and have not been experimentally
demonstrated, they compare
favorably with coherent-state BB84 schemes where $S \sim 0.1$ and a serious beam-splitting attack for 3 dB loss also obtains that
wipes out the
quantum advantage (though not the post-detection selection advantage) Bob has even with intrusion-level detection.
More significantly, Y-00 illustrates the new KCQ principle of
quantum key generation introduced in \cite{yuen04}, that creates advantage via the difference between optimal quantum receiver
performance with versus without knowledge of a secret key, which is more powerful than previous principles that rely on intrusion-level
detection.

In conclusion, the reduction of Y-00 to a classical stream cipher claimed in \cite{Nishioka} is incorrect for data bit encryption
because it still suffers from coherent-state quantum noise for typical operating parameters. It weakens both
the data and key security, possibly information-theoretically and certainly complexity-wise. It is also inapplicable to fresh key generation
because it does not recognize the seed key influence on the optimal quantum receiver performance and because it ignores privacy
amplification. The principle underlying Y-00
can be used in conjunction with additional techniques to obtain much more powerful
advantage creation for key generation, as well as near perfect information-theoretic security for the data and the key in direct
encryption against known-plaintext attacks. The detailed development has begun in \cite{yuen04} and will be presented elsewhere.

We would like to thank O. Hirota, W.Y. Hwang, and M. Ozawa for useful discussions.
This work has been supported by DARPA under grant F30602-01-2-0528.
\bibliographystyle{elsart-num}

\begin{thebibliography}{10}
\bibitem{note}
The two-mode representation in ref \cite{barbosa_1,corndorf_1,corndorf_2,corndorf_3,corndorf_4,Nishioka} is qualitatively the same as
the states (1). Quantitatively, an heterodyne measurement on (1) gives the same signal-to-noise ratio as the
``attack'' measurement in \cite{Nishioka}.
\bibitem{note2}
The notation in ref \cite{Nishioka} is somewhat different from that of \cite{barbosa_1} which we follow. Thus, our $M$ is their $2M$
and our $x_i$ is their $r_i$.
\end{thebibliography}

\end{document}